\documentclass[runningheads]{llncs}
\usepackage[style=numeric-comp, sorting=none]{biblatex}
\DeclareNameAlias{default}{family-given}
\DeclareFieldFormat{labelnumberwidth}{#1\adddot}
\usepackage{appendix}
\usepackage[labelsep=period]{caption}
\usepackage{orcidlink}
\usepackage{subcaption}
\usepackage{amsmath}
\usepackage{amsfonts}
\usepackage{algorithm}
\usepackage{algorithmic}
\usepackage{multirow}
\usepackage[T1]{fontenc}

\usepackage{graphicx}
\usepackage{subcaption}
\usepackage{booktabs}
\begin{document}
\title{Lightweight Multi-task CNN for ECG Diagnosis with GRU-Diffusion}
%
%
\author{Lehuai Xu\orcidlink{0009-0000-2951-1216} \and
Zirui Lu \and
Haoran Yang \and
Yina Zhou\thanks{Corresponding Author: \texttt{zyn@nju.edu.cn}}
}
\authorrunning{L. Xu et al.}
%
\institute{Nanjing University, Nanjing, China \\
\email{\{221180011, 221180056, 221180053\}@smail.nju.edu.cn}, 
\email{zyn@nju.edu.cn}}
\maketitle              
\begin{abstract}
With the increasing demand for real-time Electrocardiogram (ECG) classification on edge devices, existing models face challenges of high computational cost and limited accuracy on imbalanced datasets. This paper presents \textbf{Multi-task DFNet}, a lightweight multi-task framework for ECG classification across the MIT-BIH Arrhythmia Database and the PTB Diagnostic ECG Database, enabling efficient task collaboration by dynamically sharing knowledge across tasks, such as arrhythmia detection, myocardial infarction (MI) classification, and other cardiovascular abnormalities. The proposed method integrates GRU-augmented Diffusion, where the GRU is embedded within the diffusion model to capture temporal dependencies better and generate high-quality synthetic signals for imbalanced classes. The experimental results show that Multi-task DFNet achieves \textbf{99.72\%} and \textbf{99.89\%} accuracy on the MIT-BIH dataset and PTB dataset, respectively, with significantly fewer parameters compared to traditional models, making it suitable for deployment on wearable ECG monitors. This work offers a compact and efficient solution for multi-task ECG diagnosis, providing a promising potential for edge healthcare applications on resource-constrained devices.

\end{abstract}

\section{Introduction}

Electrocardiogram (ECG) signals contain rich biological information critical for the diagnosis of cardiovascular disease. Early ECG diagnostic models relied on convolutional neural networks (CNNs) \cite{han2020ml, wu2018comparison}, but these model's limited receptive fields constrained accuracy. More recent studies have focused on recurrent neural networks (RNNs) \cite{hou2019lstm, saadatnejad2019lstm, yildirim2018novel}, Transformers \cite{natarajan2020wide, yan2019fusing}, or hybrid CNN-RNN/Transformer \cite{che2021constrained, lynn2019deep, petmezas2021automated} architectures, achieving higher precision at the cost of large parameter counts and difficult training.  

On the data side, publicly available ECG datasets typically fall into two categories: arrhythmia classification and broader cardiovascular disease classification \cite{hong2020opportunities}, and most prior work addresses only one domain, with few methods capable of handling both. Moreover, samples for certain disease categories remain scarce, further degrading model performance on these underrepresented classes. Generative models have proven effective for data augmentation \cite{delaney2019synthesis, golany2019pgans}, but many studies merely adapt two-dimensional architectures to ECG, resulting in poor synthesis quality and limited augmentation benefit. These challenges hinder the application of deep learning in ECG diagnosis. At the same time, the rise of edge intelligence has created an urgent need for lightweight, efficient, and comprehensive ECG diagnostic networks.

To address these challenges, we propose a novel architecture with the following contributions:

\begin{itemize}
    \item DFNet (Dilated-Fusion Net): A lightweight one-dimensional CNN that fuses feature maps from multiple dilation rates to improve accuracy while keeping model size and computation low.  
    \item Introducing Multi-task Framework: We build a cross-dataset classification system over MIT-BIH Arrhythmia Database \cite{moody2001impact} and PTB Diagnostic ECG Database \cite{bousseljot1995nutzung} by sharing only the early DFNet encoder layers via Cross-Gating Collaboration (CGC) \cite{dai2024gated}, thereby preserving dataset-specific features while leveraging commonalities.  
    \item GRU-Diffusion Model: An enhanced diffusion generator in which gated recurrent units (GRUs) capture temporal dependencies in ECG signals, producing higher-quality synthetic samples for scarce classes and improving sensitivity to underrepresented categories.
\end{itemize}

Through these innovations, we deliver a lightweight, comprehensive, and high-accuracy ECG diagnosis model that is well-suited for energy-efficient, intelligent deployment on edge devices.

\section{Background}

\subsection{ECG signal classification}  
Early ECG diagnosis relied on handcrafted features and traditional classifiers. Deep learning, especially 1D CNNs, such as AlexNet-like architectures \cite{wu2018comparison} and ML-ResNet \cite{han2020ml}, has since improved accuracy on ECG Datasets. RNNs (LSTM, GRU) capture temporal patterns but struggle with long dependencies and imbalance \cite{hou2019lstm, lynn2019deep}. Transformers offer global attention for long-range features \cite{natarajan2020wide, romdhane2020electrocardiogram}, yet these model's high compute and memory make them ill-suited for resource-limited settings.

\subsection{Lightweight deep learning models} 
Edge deployment demands compact models. Architectures such as MobileNet \cite{howard2017mobilenets}, ShuffleNet \cite{zhang2018shufflenet}, and EfficientNet \cite{tan2019efficientnet} demonstrate efficient convolutions, but few works tailor these designs to ECG. To make it worse, such lightweight model designs may incur accuracy degradation, which is clinically unacceptable in mission-critical healthcare applications where diagnostic precision is paramount.

\subsection{Multi-task learning frameworks}
Multi-task learning shares representations across tasks to boost performance. Early shared-bottom models risk negative transfer, while Mixture of Experts or Multi-gate Mixture of Experts (MoE/MMoE) \cite{ma2018modeling, shazeer2017outrageously} improve specialization. On top of that, Cross-Gating Collaboration (CGC) \cite{dai2024gated} selectively routes features between tasks, effectively alleviating the adverse transfer effects through task-specific gating mechanisms.

\subsection{Generative models for ECG}
Data imbalance in ECG is often addressed via augmentation. Traditional signal manipulations lack diversity, and GANs \cite{goodfellow2020generative} face instability. Diffusion models are more stable and thus commonly used. Specifically, Denoising Diffusion Probabilistic Model (DDPM) \cite{ho2020denoising} formalizes generation as the reverse of a forward Markov diffusion: starting from a clean signal \(\mathbf{x}_0\), one gradually adds Gaussian noise via  
\begin{equation}
q(\mathbf{x}_t\mid \mathbf{x}_{t-1}) \;=\;\mathcal{N}\bigl(\mathbf{x}_t;\,\sqrt{\alpha_t}\,\mathbf{x}_{t-1},\,(1-\alpha_t)\,\mathbf I\bigr),
\end{equation}
so that after \(T\) steps the data becomes nearly isotropic noise. Crucially, each reverse step is also assumed to be Gaussian,  
\begin{equation}
p_\theta(\mathbf{x}_{t-1}\mid \mathbf{x}_t)\;=\;\mathcal{N}\bigl(\mathbf{x}_{t-1};\,\mathbf{\mu}_\theta(
\mathbf{x}_t,t),\,\mathbf{\Sigma}_\theta(\mathbf{x}_t,t)),
\end{equation}
where a neural network predicts the noise \(\mathbf{x}_{t-1}\) given \(\mathbf{x}_t\) and timestep \(t\). The overall generative density  
\begin{equation}
p_\theta(\mathbf{x}_0) \;=\;\int p_\theta(\mathbf{x}_0\mid \mathbf{x}_1)\,p_\theta(\mathbf{x}_1\mid \mathbf{x}_2)\cdots p_\theta(\mathbf{x}_{T-1}\mid \mathbf{x}_T)\,p(\mathbf{x}_T)\,\mathrm{d}\mathbf{x}_{1:T}
\end{equation}
is exactly the product of these reverse Markov transitions. The proposed GRU-Diffusion is built under this framework.

\vspace{-1em}
\section{Methodology}
\subsection{Overview}
\begin{figure}[H]
\centering
\includegraphics[width=\textwidth]{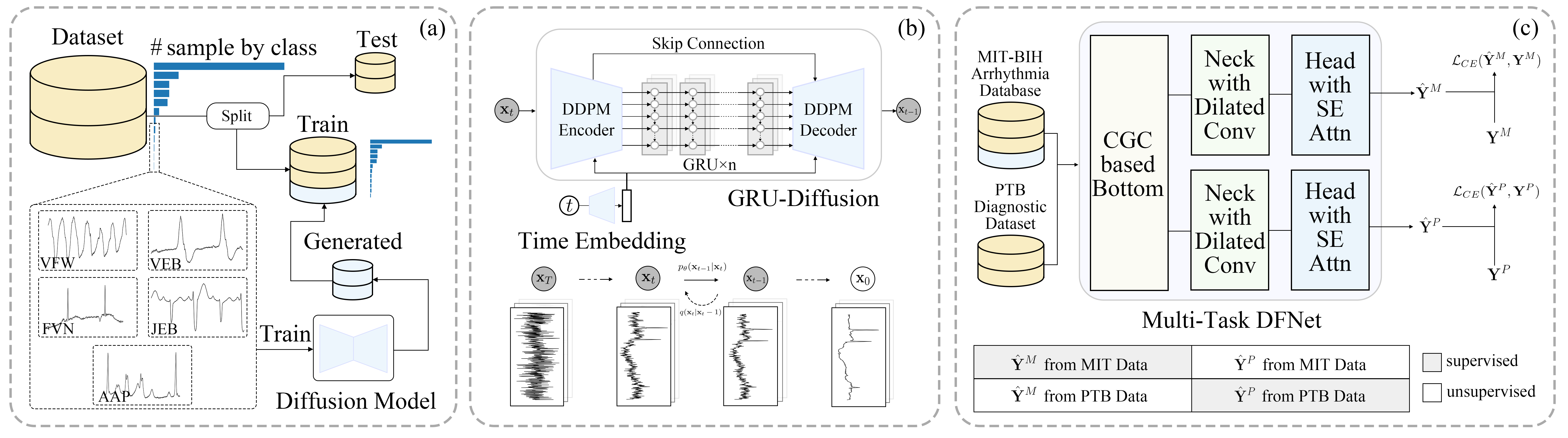}
\caption{\small Pipeline of the proposed system. (a) Dataset splitting and minority-class augmentation; (b) GRU-embedded diffusion model for ECG synthesis; (c) CGC-based multi-task DFNet for joint arrhythmia and disease classification.} \label{fig:1}
\end{figure}

We propose a two-stage architecture: (1) a generative model that enhances the dataset by supplementing rare samples. (2) a multi-task architecture that handles different ECG datasets. The architecture is illustrated in Figure \ref{fig:1}. 
\vspace{-1em}
\subsection{The Structure of GRU-Diffusion}
The structure of the proposed GRU-Diffusion is adjusted from the U-Net used in DDPM \cite{ho2020denoising}, by: (1) changing the dimension from 2D to 1D. (2) Adding multi-layer gated recurrent units (GRUs) to the middle of the U-net.

To specify, the GRU-Diffusion model contains three components: a DDPM encoder, 16 layers of GRUs, and a DDPM decoder, with time embedding to the encoder and decoder, and a skip connection between them.
\vspace{-1em}
\subsubsection{DDPM Encoder.}

Starting from the input signal \(\mathbf{X}_{signal}\in\mathbb{R}^{1\times L}\), a convolutional layer with a 7-sized kernel generates a feature map \(\mathbf{f}_0\in\mathbb{R}^{C\times L} \). The following are four encoder blocks. Each encoder block contains two residual blocks with batch normalization and SiLU activation, a multi-head self-attention layer, typically four heads, and a downsample layer, which is a convolutional layer with a stride of 2 that shortens the signal by twice but expands its channels by twice at the same time.

\vspace{-1em}
 Specifically, the downsample layer in the last encoder block is a convolutional layer with a stride of 1. It does not downsample the feature map nor change the number of channels. At the end of the encoder is another residual block. Finally, we get the output of the DDPM encoder \(\mathbf{f}_E\in\mathbb{R}^{8C\times {L\over8}}\). Typically \(C=64\). Within each encoder block, the outputs of the first residual block and the self-attention layer are extracted for skip connection.  
\vspace{-1.5em}
\subsubsection{GRU.}

The 16-layer GRU in the network follows the standard architecture of GRU \cite{chung2014empirical}, with both input size and hidden size equal to \(8C\).
\vspace{-1.5em}
\subsubsection{DDPM Decoder.}

A decoder starts with a self-attention layer and a residual block, with an output
\(\mathbf{f}_1\in\mathbb{R}^{8C\times {L\over8}}\). The following are four decoder blocks, which have the same structure as the encoder block but have upsample blocks that double the length of the feature map but decrease the number of channels in the feature map by half, instead of downsample blocks. And the two residual blocks take the concatenated product of the original feature map and the skip connection maps of the corresponding encoder block as input. Similarly, the upsample layer in the last block does not change the shape of the feature map, giving us a feature map \(\mathbf{f}_2\in\mathbb{R}^{C\times L}\). Lastly, \(\mathbf{f}_2\) goes through a final convolutional layer, producing the result \(\mathbf{f}_o\in\mathbb{R}^{1\times L}\).

\subsection{The structure of multi-task DFNet}

\begin{figure}
\centering
\includegraphics[width=0.8\textwidth]{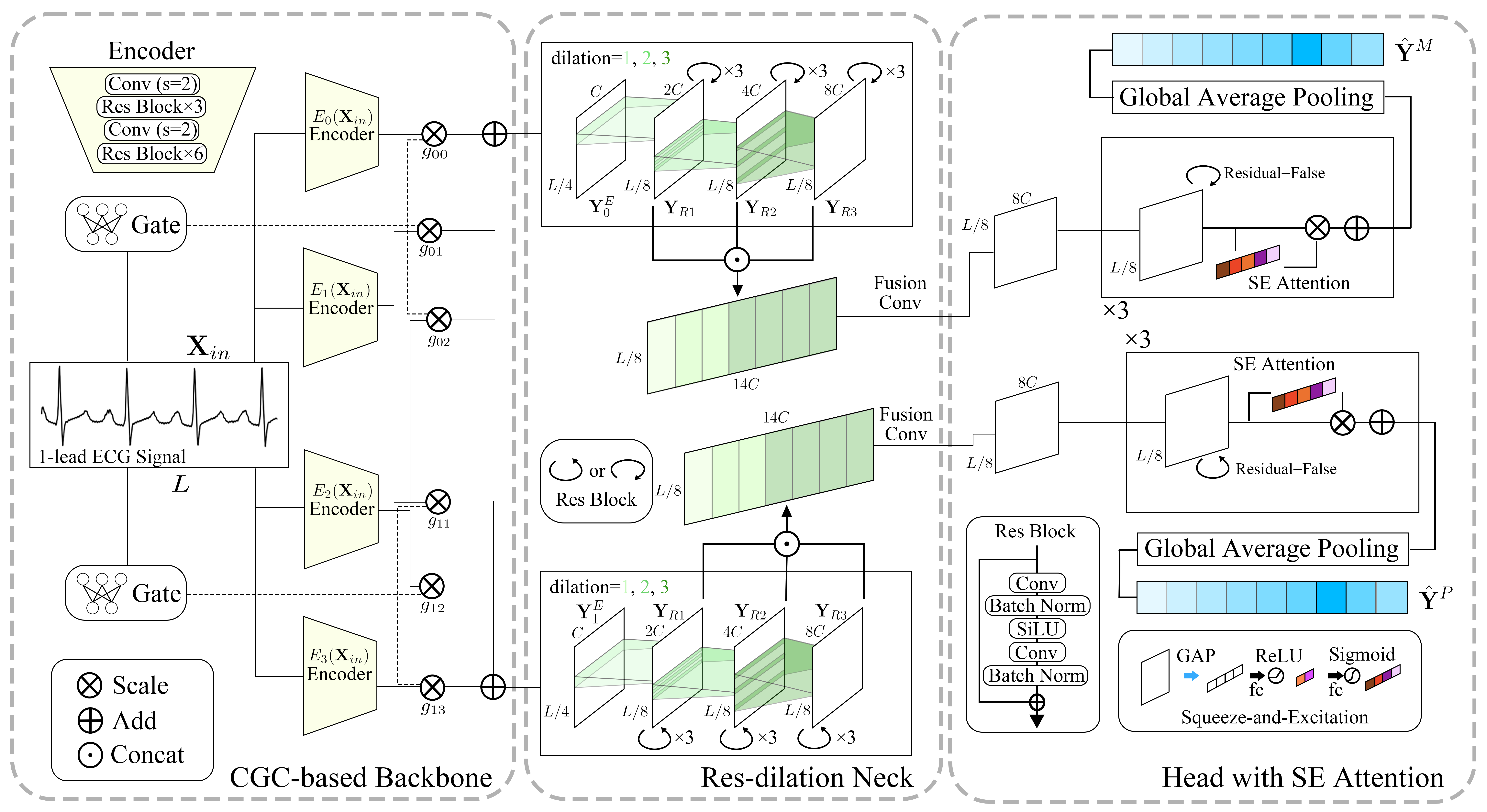}
\caption{\small Detailed architecture of Multi-task DFNet: CGC backbone, Res-dilation neck, and SE-attention head.} \label{fig:DFNet}
\end{figure}

To build the multi-task model, we first present a single-task model for classification - Dilated-Fusion Network (DFNet), which can be divided into three parts: an encoder, a neck, and a head. The thorough structure of DFNet (single-task) is shown in Table \ref{tab:DFNet}.
\vspace{-1.5em}
\subsubsection{Encoder.}

The encoder in DFNet, a traditional convolutional neural network, plays the role of shared bottom in the multi-task structure. It turns the original input \(\mathbf{X}_{in}\in\mathbb{R}^{1\times L}\) into a lower-resolution activation map \(\mathbf{X}_{act}\in\mathbb{R}^{C}\). Typically, \(C=16\). 
\vspace{-1em}
\subsubsection{Neck.}

The neck in DFNet consists of three dilated convolution blocks with \(\text{dilation rate}=1, 2, 3\) and a fusion convolutional layer. Each block includes a dilated convolutional layer and three residual blocks. The dilated convolutional layer doubles the channels in the input feature map. The first dilated convolution also downsamples the feature map by a factor of two. Let \(\mathbf{Y}_{R1}\in\mathbb{R}^{2C\times{L\over8}},\mathbf{Y}_{R2}\in\mathbb{R}^{4C\times{L\over8}},\mathbf{Y}_{R3}\in\mathbb{R}^{8C\times{L\over8}}\) denote the outputs of the three dilated convolution blocks, the production of the fusion convolutional layer, also the output of the neck, can be expressed as:
\vspace{-5pt}
\begin{equation}
\mathbf{Y}_D=\text{SiLU}(\text{BatchNorm}(\text{Conv}(\text{Concat}(\mathbf{Y}_{R1},\mathbf{Y}_{R2},\mathbf{Y}_{R3}))))
\end{equation}

where \(\mathbf{Y}_D\in\mathbb{R}^{8C\times {L\over 8}}\) is the final output.
\vspace{-1em}
\subsubsection{Head.}

The classification head in DFNet contains three residual blocks with a Squeeze-and-Excitation attention mechanism.
\vspace{-6pt}
\begin{equation}
\mathbf{Y}_{SE}=\mathbf{Y}_{in}+\underset{\text{Residual=False}}{\text{ResBlock}}(\mathbf{Y}_{in}) \cdot\sigma(F_{ex}(\text{ReLU}(F_{sq}(\text{GAP}(\mathbf{Y}_{in})))))
\end{equation}

where \( \mathbf{Y}_{SE} \in \mathbb{R}^{8C \times {L\over8}} \) is the output of the classification head, \( \mathbf{Y}_{in} \in \mathbb{R}^{8C \times {L\over8}} \) is the input feature map, and \( F_{sq} \) and \( F_{ex} \) represent the squeeze and excitation linear projections of the Squeeze-and-Excitation mechanism. \( \sigma \) is the sigmoid activation function. 
\vspace{-2pt}
The last residual block squeezes the number of channels in the map from \(8C\) to the number of classes, followed by a global average pooling layer, forming the last prediction. By global pooling, we avoid using an MLP at the end, reducing the model's parameters.
\vspace{-2em}
\begin{table}[htbp]
\centering
\caption{Layer-wise specification of the single-task DFNet backbone.}\label{tab:DFNet}
\resizebox{0.7\textwidth}{!}{
\begin{tabular}{|l|c|l|}
\hline
\textbf{Layer Name}     & \textbf{Output Size} & \textbf{Operation} \\ \hline
conv1          & $L/2$ & $\bigl[\begin{matrix}3,\,1\to8\end{matrix}\bigr],\,\mathrm{stride}=2$ \\ \hline
res1 ($\times3$)      & $L/2$ & $\bigl[\begin{matrix}k=3,\,8\to8\end{matrix}\bigr]\times2$ \\ \hline
conv2          & $L/4$ & $\bigl[\begin{matrix}3,\,8\to16\end{matrix}\bigr],\,\mathrm{stride}=2$ \\ \hline
res2 ($\times6$)      & $L/4$ & $\bigl[\begin{matrix}3,\,16\to16\end{matrix}\bigr]\times2$ \\ \hline
conv3          & $L/8$  & $\bigl[\begin{matrix}3,\,16\to32\end{matrix}\bigr],\,\mathrm{stride}=2$ \\ \hline
res3 ($\times3$)      & $L/8$  & $\bigl[\begin{matrix}3,\,32\to32\end{matrix}\bigr]\times2$ \\ \hline
conv4          & $L/8$  & $\bigl[\begin{matrix}3,\,32\to64\end{matrix}\bigr],\,\mathrm{dilation}=2$ \\ \hline
res4 ($\times3$)      & $L/8$  & $\bigl[\begin{matrix}3,\,64\to64\end{matrix}\bigr]\times2$ \\ \hline
conv5          & $L/8$  & $\bigl[\begin{matrix}3,\,64\to128\end{matrix}\bigr],\,\mathrm{dilation}=3$ \\ \hline
res5 ($\times3$)      & $L/8$  & $\bigl[\begin{matrix}3,\,128\to128\end{matrix}\bigr]\times2$ \\ \hline
fusion\_conv   & $L/8$  & $\bigl[\begin{matrix}3,\,224\to128\end{matrix}\bigr]$ \\ \hline
SE-ResBlock ($\times2$) & $L/8$ & $[\,\mathrm{SE},\,3,\,128\to128\,]\times2$ \\ \hline
SE-ResBlock    & $L/8$  & $[\,\mathrm{SE},\,3,\,128\to \text{cls\_num}\,]$ \\ \hline
pooling        & 1   & Adaptive Avg Pooling \\ \hline
\end{tabular}}
\end{table}

\subsection{The multi-task architecture}

The multi-task model is built based on the DFNet Network. The schematic diagram of multi-task DFNet is shown in Figure \ref{fig:DFNet}.
\vspace{-1em}
\subsubsection{CGC-based Backbone.} 

The backbone utilizes the CGC framework with four encoders of DFNet, where the output of the encoder modules is aggregated to form the final feature representation. Let \(\mathbf{X}_{in}\) denote the input signal, we have:
  
\begin{equation}
\begin{bmatrix}
\mathbf{Y}^E_0\\
\mathbf{Y}^E_1
\end{bmatrix}
=
\begin{bmatrix}
g_{00} & g_{01} & g_{02} & 0 \\
0 & g_{11} & g_{12} & g_{13}
\end{bmatrix}
\begin{bmatrix}
E_0(\mathbf{X}_{in}) \\
E_1(\mathbf{X}_{in}) \\
E_2(\mathbf{X}_{in}) \\
E_3(\mathbf{X}_{in}) \\
\end{bmatrix}
\end{equation}

where \(\mathbf{Y}^E_i\) is the aggregated feature for the \(i\)-th task \(i\in[0, 1]\), \(N\) is the number of encoders, \( g_{ik} \) are the weights controlling the contribution of each encoder, which is linear projected from the input signal, and \( E_k(\mathbf{X}) \) denotes the output of the \( k \)-th encoder.
\vspace{-1em}
\subsubsection{Multi-task tower.} 

The multi-task tower consists of a neck and a head in DFNet. Different towers are independent of each other.
\vspace{-1em}
\subsubsection{Loss function and training strategy.}

The loss function of the multi-task DFNet comprises three parts: cross-entropy losses of the two tasks, and \(\mathcal{L}_2\) regularization loss. It is given by:

\begin{equation}
\begin{aligned}
\mathcal{L}_\text{total} &= \alpha\mathcal{L}_\text{CE}(\hat{\mathbf{Y}}^M,\mathbf{Y}^M)+\beta\mathcal{L}_\text{CE}(\hat{\mathbf{Y}}^P,\mathbf{Y}^P)+{\lambda\over2}\mathcal{L}_2 \\
&=-\alpha \sum_{i=1}^{N} \mathbf{Y}^M_i \log(\hat{\mathbf{Y}}^M_i)
   - \beta \sum_{i=1}^{N} \mathbf{Y}^P_i \log(\hat{\mathbf{Y}}^P_i)
   + \frac{\lambda}{2} \sum_{i} \| \mathbf{W}_i \|_2^2
\end{aligned}
\end{equation}

where \(\alpha\), \(\beta\) and \(\lambda\) are the weights of different losses. \(\hat{\mathbf{Y}}^M\) and \(\mathbf{Y}^M\) are the prediction and target of task \(M\), while \(\hat{\mathbf{Y}}^P\) and \(\mathbf{Y}^P\) are the prediction and target of task \(P\). \(\mathbf{W}_i\) represent the weights of the model.

In contrast to conventional multi-task learning frameworks that typically employ multiple annotations within a single dataset, the current approach involves distinct datasets containing annotations of separate classes. Consequently, a half-blind training strategy is employed to optimize model performance. The proposed training strategy is comprehensively outlined in Algorithm \ref{alg:training}.

\begin{algorithm}
\caption{Multi-task Training Strategy}
\label{alg:training}
\begin{algorithmic}[1]
\STATE \textbf{Input:} Two datasets, with batch size \(n_B\): 
\begin{itemize}
    \item \(\mathcal{D}^M = \bigl\{(\mathbf{X}^M_i, \mathbf{Y}^M_i)\mid 
      \mathbf{X}^M_i \in \mathbb{R}^{C\times L},\;
      \mathbf{Y}^M_i\in\{1,\dots,K_M\}
    \bigr\}\), dataset 1.
    \item \(\mathcal{D}^P = \bigl\{(\mathbf{X}^P_j, \mathbf{Y}^P_j)\mid 
      \mathbf{X}^P_j \in \mathbb{R}^{C\times L},\;
      \mathbf{Y}^P_j\in\{1,\dots,K_P\}
    \bigr\}\), dataset 2.
\end{itemize}
\STATE Initialize the model \(\mathcal{M}\).
\FOR{\textbf{each epoch} \(e = 1, 2, \dots, E\)}
    \FOR{step = 1 to max(len(\(\mathcal{X}^M\)), len(\(\mathcal{X}^P\)))}
        \STATE Sample a batch of \(\mathbf{X}^M,\mathbf{Y}^M\) from \(\mathcal{D}^M\), and \(\mathbf{X}^P,\mathbf{Y}^P\) from \(\mathcal{D}^P\).
        \STATE Forward: \(\hat{\mathbf{Y}}^m, \hat{\mathbf{Y}}^p = \mathcal{M}(\texttt{Concat}(\mathbf{X}^M, \mathbf{X}^P))\)
        \STATE Extract prediction: \(\hat{\mathbf{Y}}^M=\hat{\mathbf{Y}}^m[0{:}n_B], \hat{\mathbf{Y}}^P=\hat{\mathbf{Y}}^p[n_B{:}2n_B]\)
        \STATE Compute the loss:
        \(
        \mathcal{L}_\text{total} = \alpha\mathcal{L}_\text{CE}(\hat{\mathbf{Y}}^M,\mathbf{Y}^M)+\beta\mathcal{L}_\text{CE}(\hat{\mathbf{Y}}^P,\mathbf{Y}^P)+{\lambda\over2}\mathcal{L}_2
        \)
        \STATE Perform backpropagation and update the model parameters.
    \ENDFOR
\ENDFOR
\end{algorithmic}
\end{algorithm}

\vspace{-2em}

\section{Experiment}

\subsection{Environment Setup}

All experiments are conducted on a machine equipped with an NVIDIA RTX 3090 GPU and an Intel Xeon Gold 6330 CPU. The model is trained using the Adam optimizer with a learning rate of \(1 \times 10^{-3}\) and an L2 regularization coefficient (\(\lambda\)) of \(10^{-5}\). A batch size of 1000 is utilized, and training is carried out over 300 epochs. The ReduceLROnPlateau scheduler(mode="min", factor=0.5, patience=25) is employed to adjust the learning rate dynamically.

\subsection{Dataset Description}

\begin{figure}[htbp]
  \centering
  \begin{subfigure}[b]{0.4\linewidth}
    \centering
    \includegraphics[width=0.88\linewidth, height=0.76\linewidth]{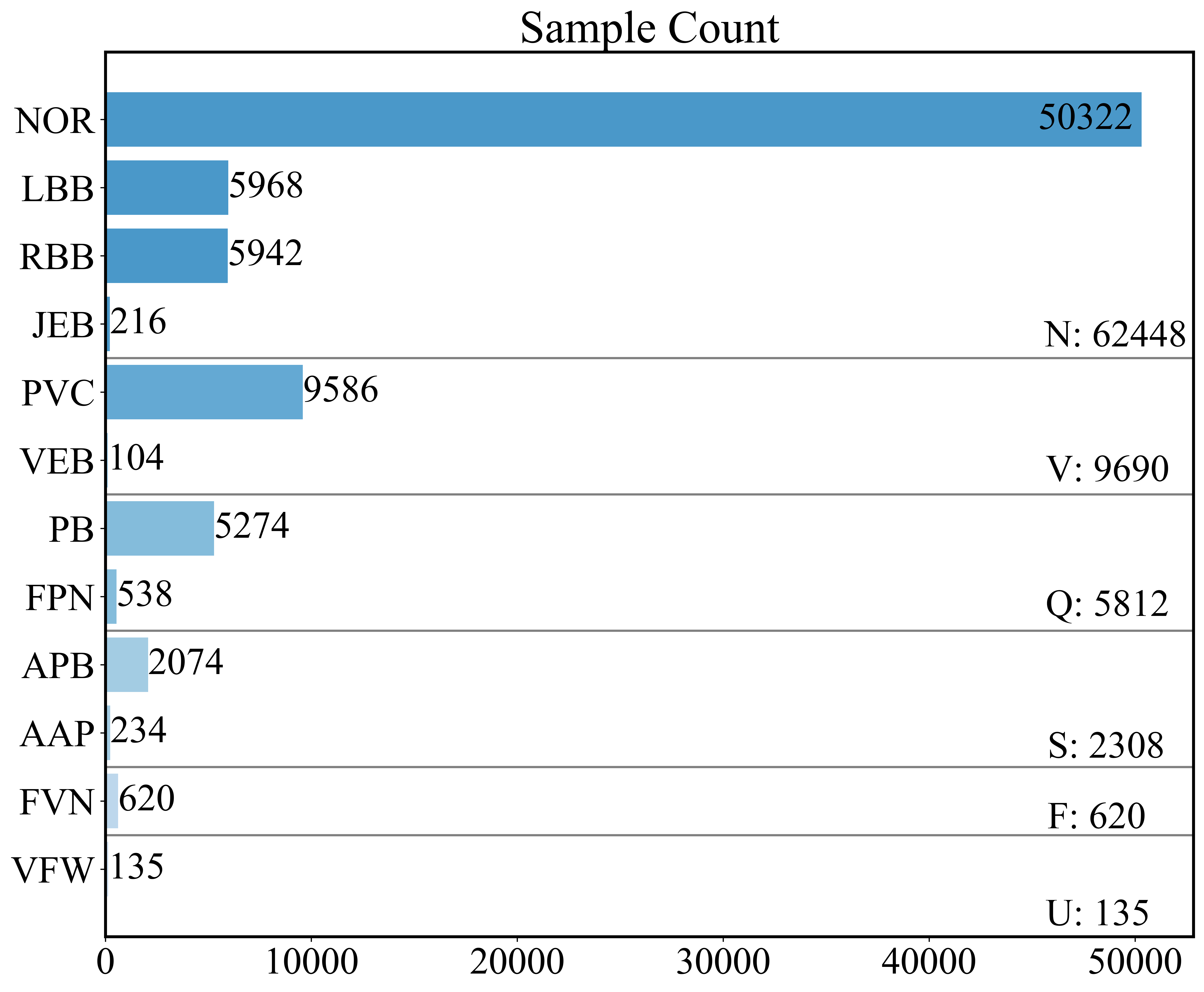}
    \caption*{(a) MIT-BIH}
  \end{subfigure}
  \begin{subfigure}[b]{0.4\linewidth}
    \centering
    \includegraphics[width=0.88\linewidth, height=0.76\linewidth]{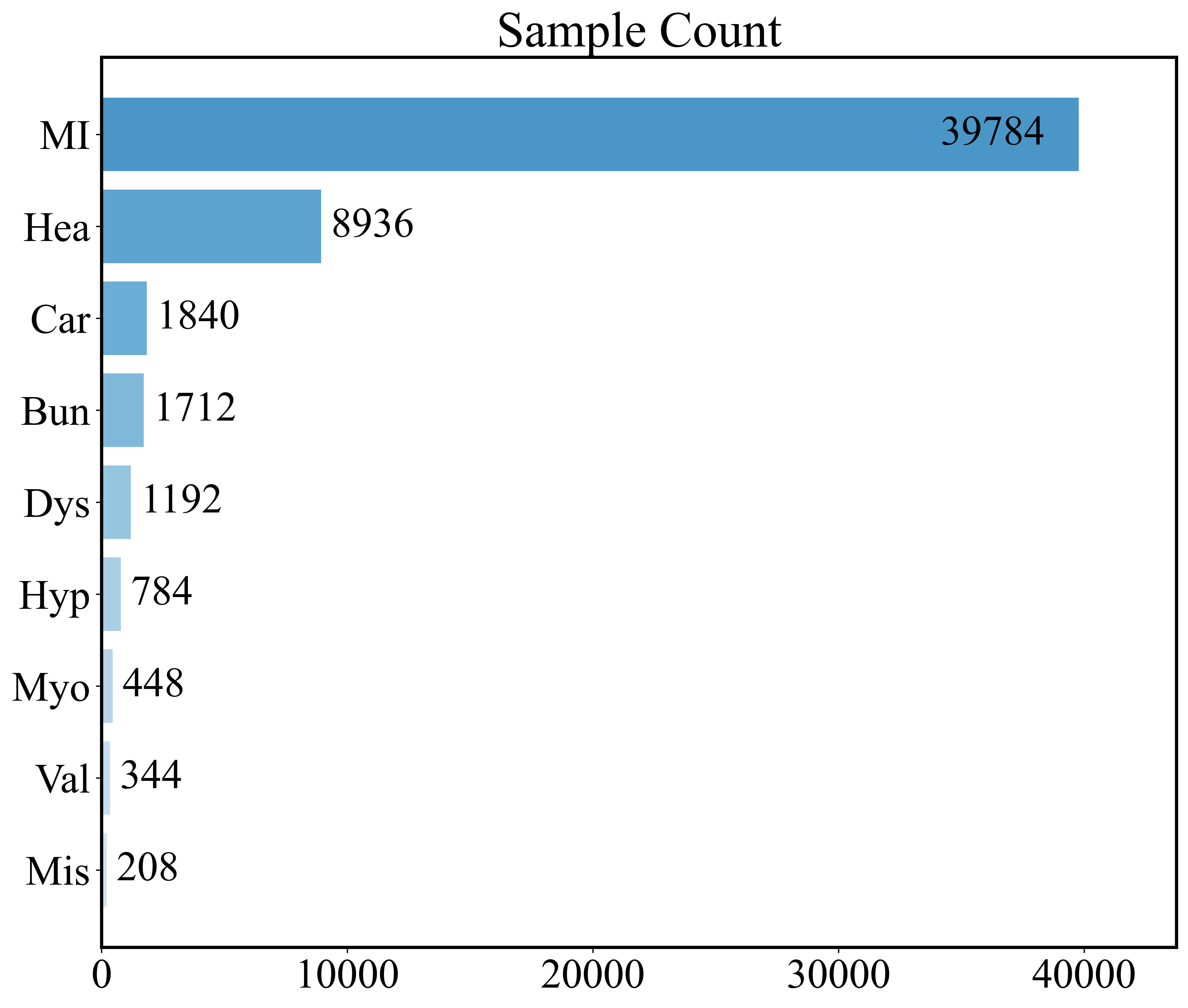}
    \caption*{(b) PTB}
  \end{subfigure}

  \caption{Class-wise sample distribution in (a) MIT-BIH and (b) PTB before augmentation.}
  \label{fig:sample}
\end{figure}

The proposed network is evaluated on the MIT-BIH Arrhythmia Database \cite{moody2001impact} and the PTB Diagnostic ECG Database \cite{bousseljot1995nutzung}. We selected 12 out of the 16 available classes from the 2-lead MIT-BIH dataset, and 9 classes from the 15-lead PTB dataset (we only use the standard 12 leads). All data was resampled to 250 Hz and split into single-lead signals of 512 sample points. Across both datasets, 80\% of the data were used for training and 20\% for testing (shuffled). The total number of samples is shown in Figure \ref{fig:sample}.

In the dataset of MIT-BIH, the different heartbeat classes are as follows: Normal beat (NOR), Left bundle branch block beat (LBB), Paced beat (PB), Fusion of paced and normal beat (FPN), Right bundle branch block beat (RBB), Atrial premature beat (APB), Aberrated atrial premature beat (AAP), Premature ventricular contraction (PVC), Fusion of ventricular and normal beat (FVN), Ventricular flutter wave (VFW), Nodal (junctional) escape beat (JEB), and Ventricular escape beat (VEB). To address the class imbalance, we used the GRU-Diffusion model to augment the five minority classes (AAP, FVN, VFW, JEB, and VEB) by generating 900 additional samples each.

The PTB dataset consists of the following classes: Bundle branch block (Bun), Cardiomyopathy (Car), Dysrhythmia (Dys), Hypertrophy (Hyp), Miscellaneous (Mis), Myocardial infarction (MI), Healthy controls (Hea), Myocarditis (Myo) and Valvular (Val) heart disease.
\vspace{-0.5em}
\subsection{Quantitative and Qualitative Evaluation on GRU-Diffusion}
\vspace{-0.5em}

\begin{figure}[t]
  \centering

  \begin{subfigure}[b]{0.32\linewidth}
    \centering
    \includegraphics[width=\linewidth]{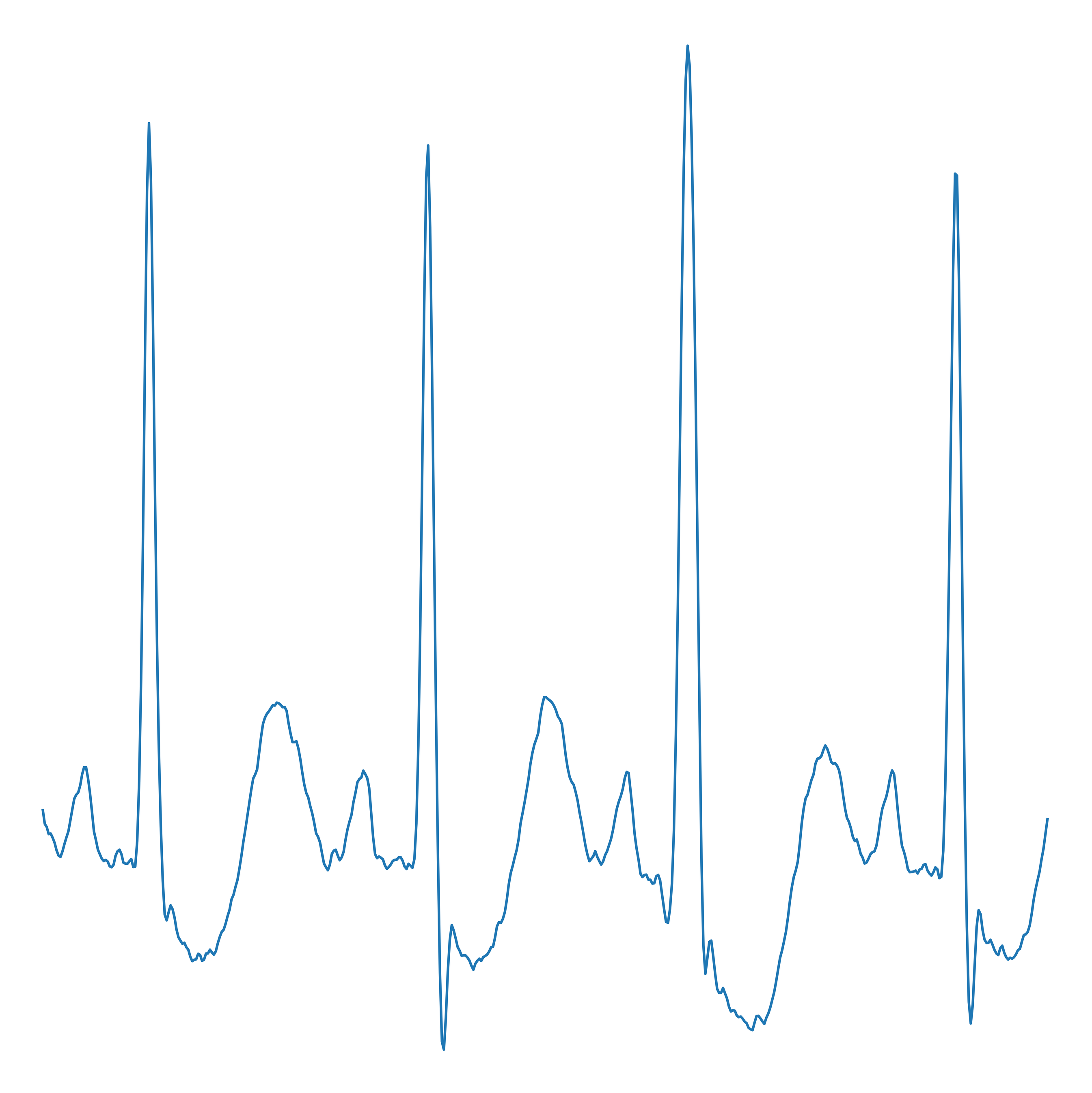}
  \end{subfigure}
  \begin{subfigure}[b]{0.32\linewidth}
    \centering
    \includegraphics[width=\linewidth]{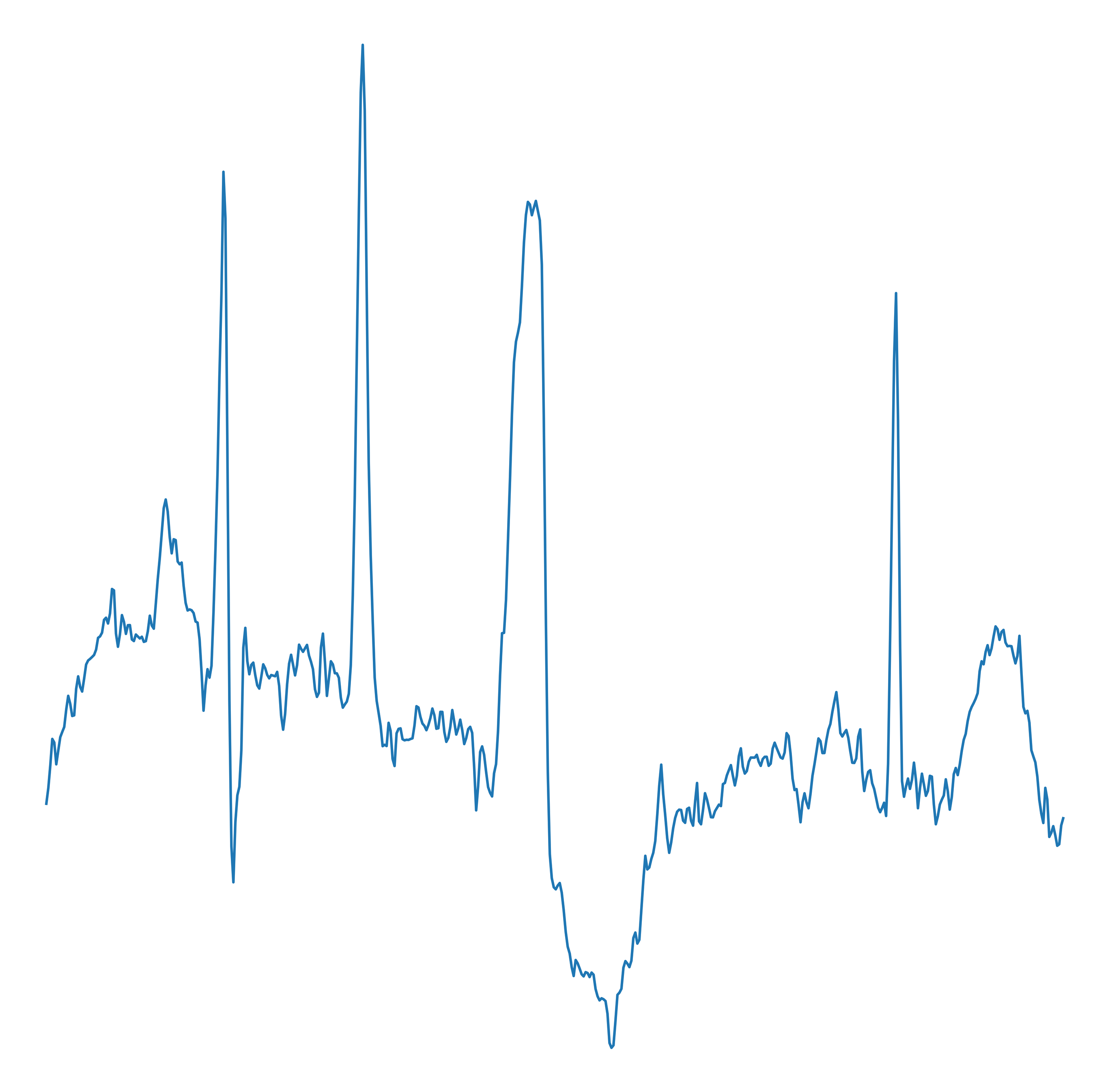}
  \end{subfigure}
  \begin{subfigure}[b]{0.32\linewidth}
    \centering
    \includegraphics[width=\linewidth]{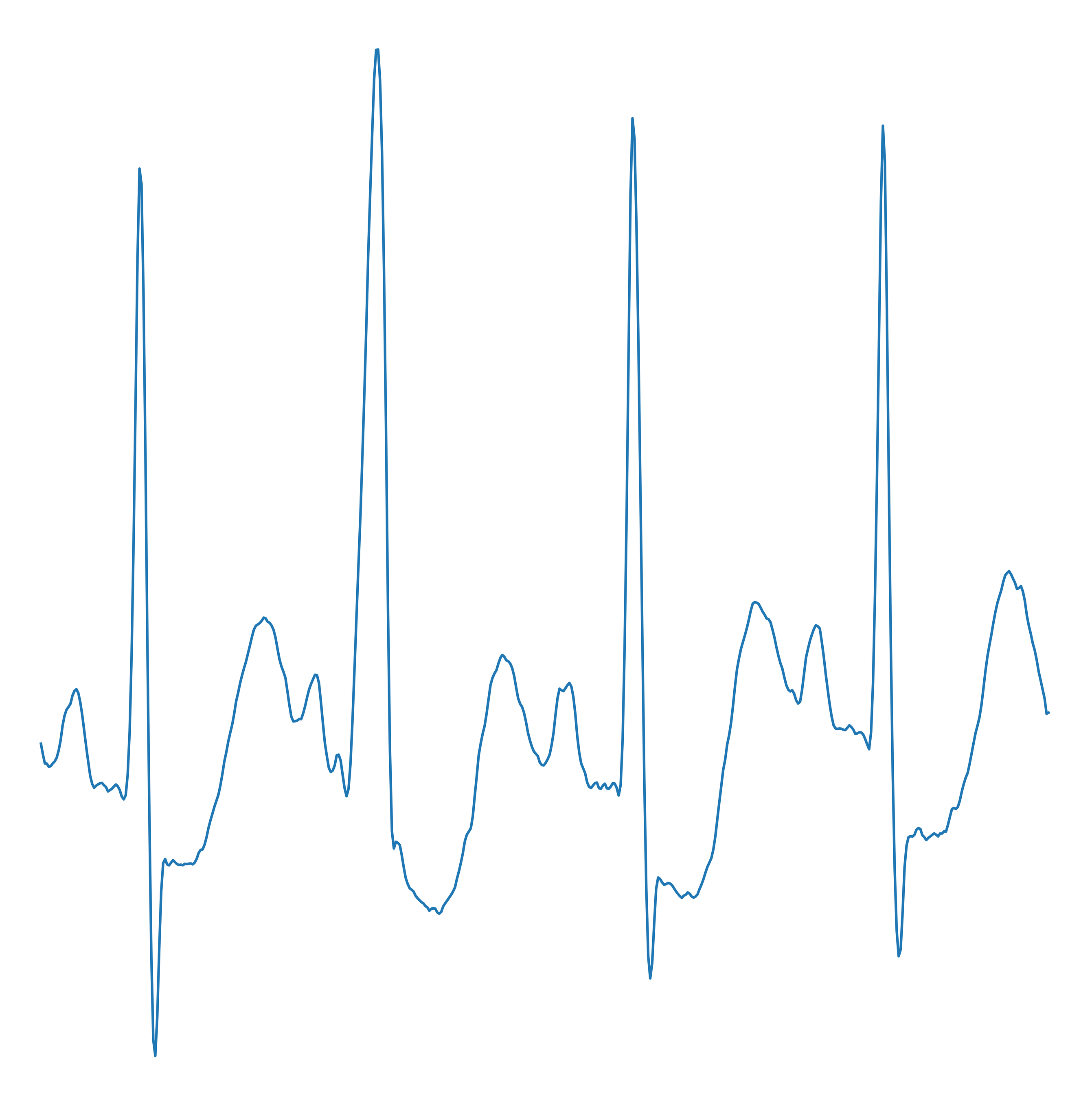}
  \end{subfigure}

  \begin{subfigure}[b]{0.32\linewidth}
    \centering
    \includegraphics[width=\linewidth]{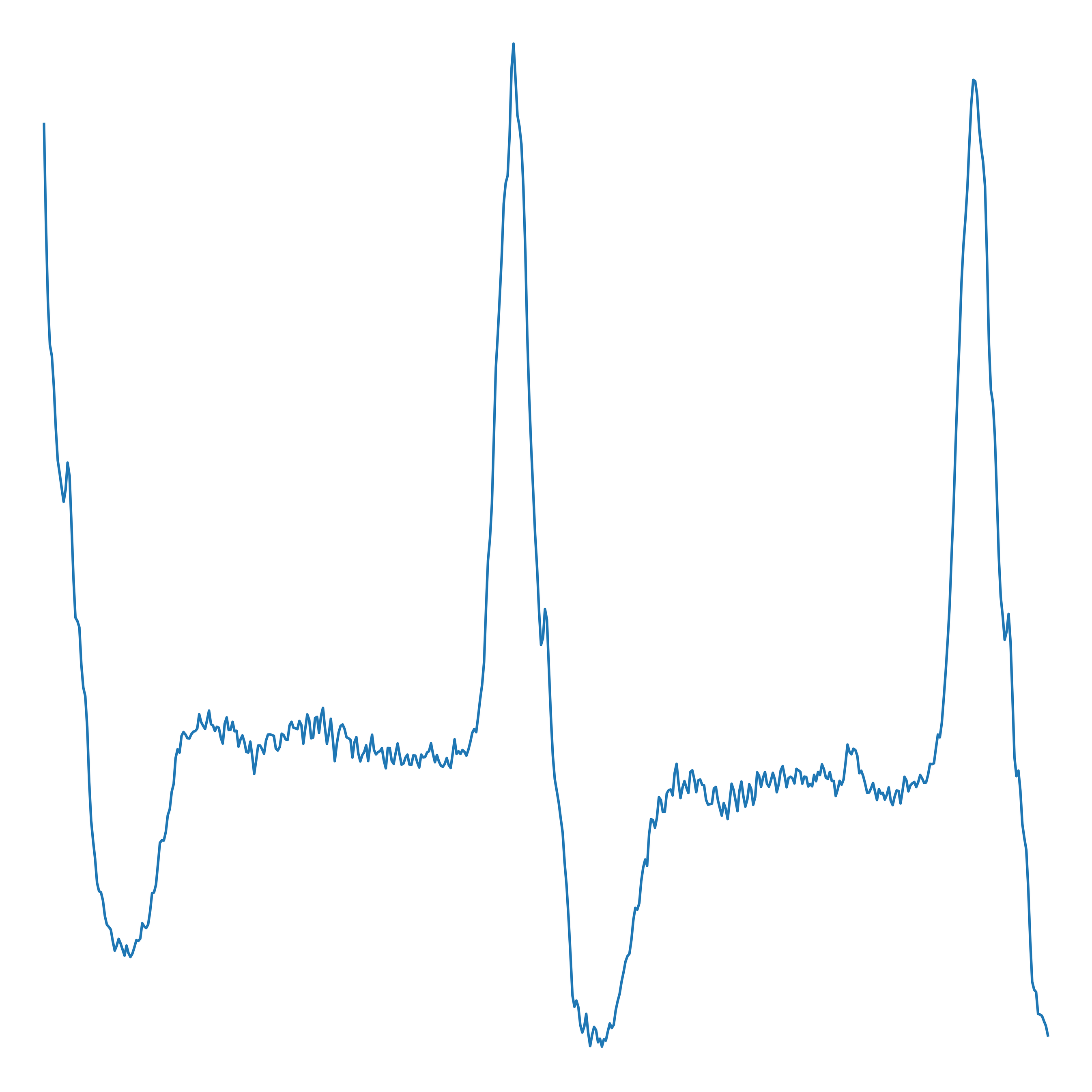}
    \caption*{True}
  \end{subfigure}
  \begin{subfigure}[b]{0.32\linewidth}
    \centering
    \includegraphics[width=\linewidth]{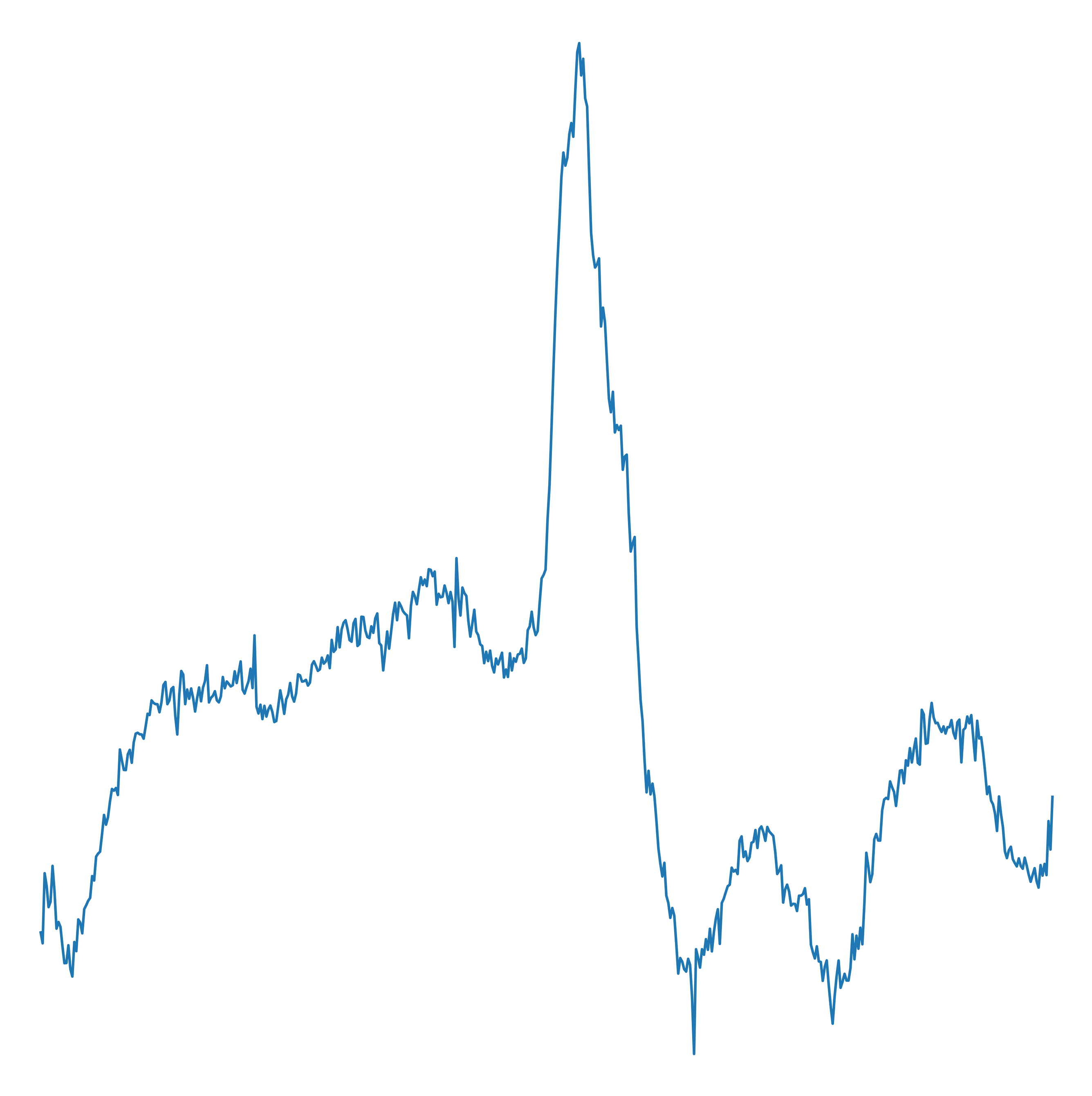}
    \caption*{DDPM}
  \end{subfigure}
  \begin{subfigure}[b]{0.32\linewidth}
    \centering
    \includegraphics[width=\linewidth]{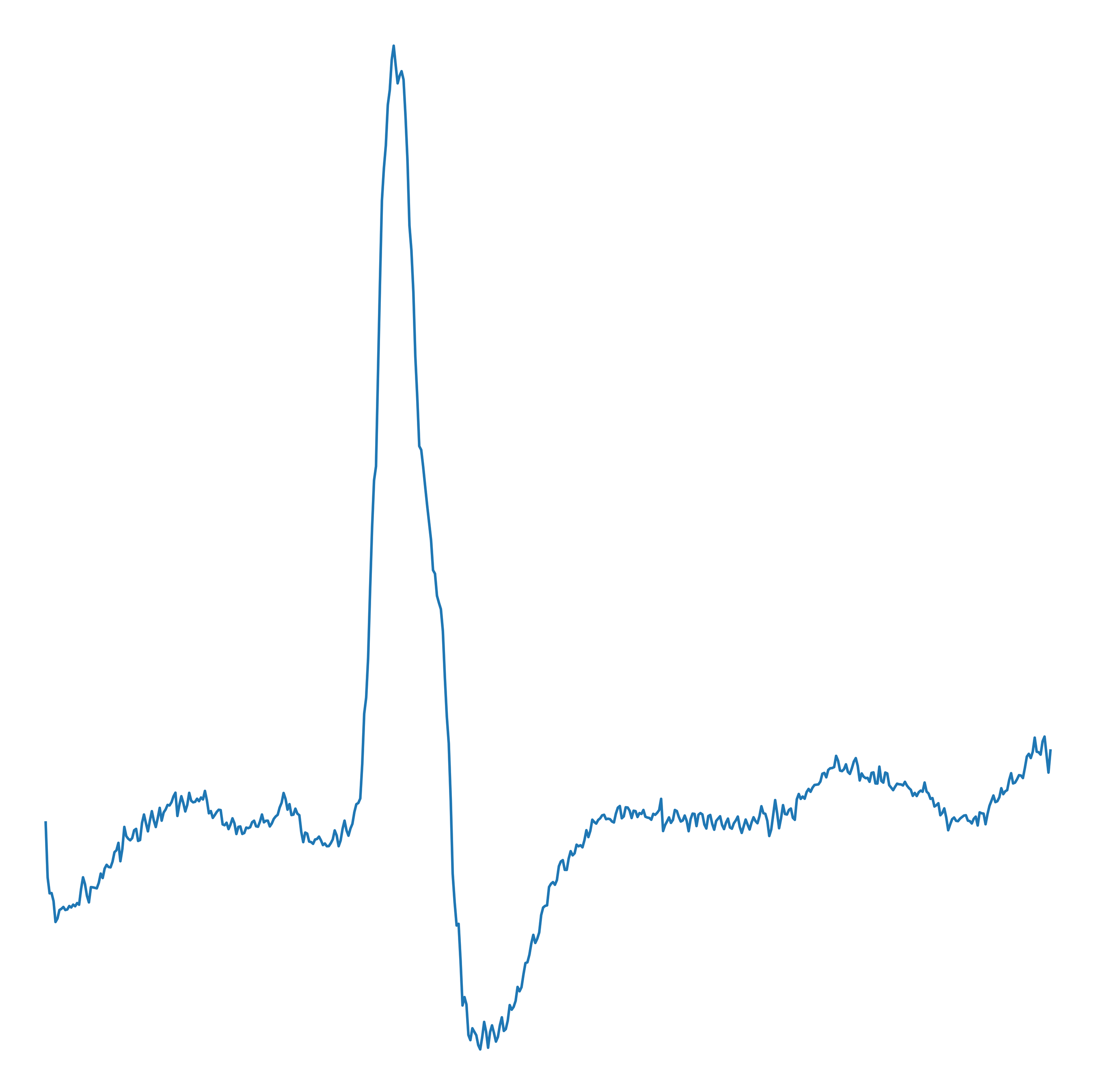}
    \caption*{GRU-Diffusion}
  \end{subfigure}

  \caption{Visual comparison of real beats and beats generated by DDPM and GRU-Diffusion.}
  \label{fig:plot}
\end{figure}

\begin{table}[t]
    \centering
    \captionof{table}{\small Quality metrics for DDPM vs. GRU-Diffusion on five rare MIT-BIH classes.}\label{ddpm}
    \resizebox{\textwidth}{!}{
    \begin{tabular}{l|llll|llll}
    \hline
      & \multicolumn{4}{|l|}{DDPM}                                                                                                                       & \multicolumn{4}{l}{GRU-Diffusion}                                                                                                                                \\ \cline{2-9} 
      & \multicolumn{1}{l|}{FID$\downarrow$}     & \multicolumn{1}{l|}{$\mu_{DTW}\downarrow$} & \multicolumn{1}{l|}{$\sigma_{DTW}\downarrow$} & KL$\downarrow$    & \multicolumn{1}{l|}{FID$\downarrow$}             & \multicolumn{1}{l|}{$\mu_{DTW}\downarrow$} & \multicolumn{1}{l|}{$\sigma_{DTW}\downarrow$} & KL$\downarrow$             \\ \hline

    AAP & \multicolumn{1}{l|}{5.2751}   & \multicolumn{1}{l|}{58.4084}  & \multicolumn{1}{l|}{37.3638}  & 0.0348 & \multicolumn{1}{l|}{4.3648}   & \multicolumn{1}{l|}{56.8948}  & \multicolumn{1}{l|}{25.6112} & 0.0463 \\

    FVN & \multicolumn{1}{l|}{93.3865}  & \multicolumn{1}{l|}{165.4199} & \multicolumn{1}{l|}{70.7186}  & 0.0228 & \multicolumn{1}{l|}{28.7431}  & \multicolumn{1}{l|}{163.1977} & \multicolumn{1}{l|}{54.4970} & 0.0226 \\

    VFW & \multicolumn{1}{l|}{137.1146} & \multicolumn{1}{l|}{220.9558} & \multicolumn{1}{l|}{114.6480} & 0.0504 & \multicolumn{1}{l|}{22.8891}  & \multicolumn{1}{l|}{141.6840} & \multicolumn{1}{l|}{83.3020} & 0.0134 \\

    JEB & \multicolumn{1}{l|}{31.8519}  & \multicolumn{1}{l|}{95.4929}  & \multicolumn{1}{l|}{63.9846}  & 0.0726 & \multicolumn{1}{l|}{4.7920}   & \multicolumn{1}{l|}{74.7268}  & \multicolumn{1}{l|}{59.8243} & 0.0153 \\

    VEB & \multicolumn{1}{l|}{54.5154}  & \multicolumn{1}{l|}{126.0979} & \multicolumn{1}{l|}{111.9180} & 0.0401 & \multicolumn{1}{l|}{19.8446}  & \multicolumn{1}{l|}{118.4965} & \multicolumn{1}{l|}{88.3082} & 0.0152 \\ \hline

    Average & \multicolumn{1}{l|}{64.4287} & \multicolumn{1}{l|}{133.2750} & \multicolumn{1}{l|}{79.7266} & 0.0441 & \multicolumn{1}{l|}{\textbf{16.1267}} & \multicolumn{1}{l|}{\textbf{111.0000}} & \multicolumn{1}{l|}{\textbf{62.3085}} & \textbf{0.0226} \\ \hline

    \end{tabular}}
\end{table}

We conducted quantitative and qualitative assessments using multiple metrics to evaluate the quality of synthetic ECG signals generated by the GRU-Diffusion model. Specifically, we applied Principal Component Analysis (PCA) to reduce the dimensionality of real and generated signals, followed by the calculation of the \textbf{Frechet Distance (FID)} to assess distribution similarity. Furthermore, we computed the average \textbf{Dynamic Time Warping (DTW)} distance between randomly paired real and generated signals to evaluate temporal alignment. To analyze the frequency characteristics, we estimated the average \textbf{Power Spectral Density (PSD)} of signals using the Welch method. We calculated the \textbf{Kullback-Leibler (KL) divergence} between the PSDs of real and generated signals. 

Quantitative results, summarized in Table \ref{ddpm}, demonstrate that GRU-Diffusion achieves lower FID, DTW mean (\(\mu_{\text{DTW}}\)), and KL divergence compared to the baseline DDPM model, indicating improved fidelity and temporal consistency. Additionally, Figure \ref{fig:plot} visualizes examples of real signals alongside those generated by DDPM and GRU-Diffusion, highlighting the superior morphological quality achieved by GRU-Diffusion.

\subsection{Quantitative Evaluation on DFNet}

We compare the parameter efficiency, FLOPs, and F1 score of the proposed DFNet with several baseline models, including classic architectures such as VGG \cite{simonyan2014very}, AlexNet \cite{krizhevsky2017imagenet}, ResNet \cite{he2016deep}, and SENet \cite{hu2018squeeze}, as well as lightweight models like MobileNet \cite{howard2017mobilenets}, ShuffleNet \cite{zhang2018shufflenet}, and EfficientNet \cite{tan2019efficientnet} in Table \ref{baseline}. The results demonstrate that DFNet achieves superior classification performance while significantly reducing model size and computational cost compared to traditional and lightweight baselines, making it highly suitable for real-time edge-device applications.
\vspace{-2em}
\begin{table}[]
    \centering
    \caption{\small Efficiency comparison: parameters, FLOPs and 12-class F1 of DFNet (single/multi) vs. baseline CNNs.}\label{baseline}
    \resizebox{0.75\textwidth}{!}{
    \begin{tabular}{l|l|l|l|l}
    \hline
    models(1D)            & parameters       & GFLOPS        & F1 on MITBIH(\%) &  F1 on PTB(\%)  \\ \hline
    ResNet-18             & 3.85M            & 0.09          & 92.33            & 98.43 \\
    ResNet-34             & 7.22M            & 0.18          & 92.78            & 98.72 \\
    ResNet-50             & 15.98M           & 0.40          & 92.88            & 98.50 \\
    SE-ResNet-18          & 3.94M            & 0.09          & 93.70            & 98.47 \\
    SE-ResNet-34          & 7.38M            & 0.18          & 94.01            & 98.80 \\
    SE-ResNet-50          & 18.49M           & 0.40          & 94.14            & 98.53 \\
    AlexNet               & 23.90M           & 0.05          & 90.57            & 98.03 \\
    VGG-16                & 36.42M           & 0.32          & 93.00            & 98.59 \\
    VGG-19                & 38.19M           & 0.42          & 93.03            & 98.74 \\ \hline
    MobileNetV2           & 1.15M            & \textbf{0.02} & 92.63            & 98.25 \\
    MobileNetV3           & 1.45M            & 0.03          & 93.88            & 98.32 \\
    ShuffleNetV2          & 1.71M            & 0.06          & 91.85            & 98.68 \\
    EfficientNetB0        & 3.40M            & 0.07          & 93.86            & 98.61 \\
    \hline
    \textbf{DFNet}        & \textbf{529.05K} & 0.03          & 94.22            & 98.83 \\
    \textbf{Multi-Task DFNet}  & 1.07M       & 0.07          & \textbf{95.26}   & \textbf{99.09} \\ \hline
    \end{tabular}}
\end{table}

\vspace{-2em}
\subsubsection{Ablation Study of DFNet.}

We conducted an ablation study to evaluate the contributions of key components in the DFNet model, including the fusion layer, SE Attention, and dilated convolutions. As shown in Table \ref{ablation}, removing any of these components led to a performance drop, with the absence of the dilated convolutions causing the most significant decline in F1 score. 
\vspace{-2em}
\begin{table}[]
\centering
\caption{Ablation on DFNet components (fusion, SE, dilation)}\label{ablation}
\resizebox{0.6\textwidth}{!}{
\begin{tabular}{l|llll|llll}
\hline
 & \multicolumn{4}{c|}{MIT-BIH} & \multicolumn{4}{c}{PTB} \\ \cline{2-9}
 & Acc(\%) & Prec(\%) & Rec(\%) & F1(\%) & Acc(\%) & Prec(\%) & Rec(\%) & F1(\%) \\ \hline

No Fusion    & 98.48 & 96.63 & 90.39 & 93.20 & 99.73 & 98.82 & 98.52 & 98.67 \\
No SE Attn   & 98.20 & 96.89 & 89.42 & 92.81 & 99.65 & 99.59 & 98.33 & 98.46 \\
No Dilation  & 98.30 & 96.37 & 90.33 & 93.05 & 99.76 & 98.65 & 98.29 & 98.46 \\ \hline

\textbf{Full model} &
\textbf{99.01} & \textbf{98.02} & \textbf{91.11} & \textbf{94.22} &
\textbf{99.86} & \textbf{99.09} & \textbf{98.62} & \textbf{98.83} \\ \hline

\end{tabular}}
\end{table}

\vspace{-3em}
\subsection{Quantitative Evaluation on the complete framework}

\subsubsection{Overall Performance Comparison.}
We compare the classification performance of the complete framework, Multi-task DFNet + GRU-Diffusion, against several different methods using accuracy (acc), precision (prec), recall (rec), and F1 score, as shown in Table \ref{comparisonwithothers}. To align with previous works, we reorganized the 12-class classification results on the MIT-BIH dataset according to the AAMI standard into six major classes when calculating these metrics. See Appendix Table \ref{bigtab} for full details. 

\vspace{-1em}
\begin{table}[htbp]
\centering
\caption{\small Performance of Multi-task DFNet against recent ECG classifiers on MIT-BIH (6-class AAMI metrics) and PTB}\label{comparisonwithothers}
\begin{minipage}[t]{0.48\textwidth}
  \centering
  \resizebox{\textwidth}{!}{
  \begin{tabular}{l|llll}
    \hline
    Methods (MIT-BIH) & Acc(\%) & Prec(\%) & Rec(\%) & F1(\%) \\ \hline
    Dang et al.~\cite{dang2019novel} & 95.48 & 96.53 & 87.74 & 91.90 \\
    Li et al.~\cite{li2019automated} & 99.50 & 97.30 & 98.10 & 97.70 \\
    Shaker et al.~\cite{shaker2020generalization} & 98.00 & 90.00 & 97.70 & 93.70 \\
    Ahmad et al.~\cite{ahmad2021ecg} & 99.70 & 98.00 & 98.00 & 98.00 \\
    Taissir F. Romdhane et al.~\cite{romdhane2020electrocardiogram} & 98.47 & 98.43 & 98.47 & 98.43 \\ 
    \hline
    \textbf{Proposed Multitask-DFNet} & \textbf{99.76} & \textbf{99.73} & \textbf{98.74} & \textbf{99.23} \\ \hline
  \end{tabular}}
\end{minipage}
\hfill
\begin{minipage}[t]{0.48\textwidth}
  \centering
  \resizebox{\textwidth}{!}{
  \begin{tabular}{l|llll}
    \hline
    Methods (PTB) & Acc(\%) & Prec(\%) & Rec(\%) & F1(\%) \\ \hline
    Acharya et al.~\cite{acharya2017application} & 95.22 & 95.49 & 94.19 & 94.80 \\
    Kachuee et al.~\cite{kachuee2018ecg} & 95.90 & 95.20 & 95.10 & 95.20 \\
    Liu et al.~\cite{liu2017real} & 96.00 & 97.37 & 95.40 & 96.30 \\
    Sharma et al.~\cite{sharma2015multiscale} & 96.00 & 99.00 & 93.00 & 95.90 \\
    Ahmad et al.~\cite{ahmad2021ecg} & 99.20 & 98.00 & 98.00 & 98.00 \\ \hline
    \textbf{Proposed Multitask-DFNet} & \textbf{99.89} & \textbf{99.24} & \textbf{99.13} & \textbf{99.18} \\ \hline
  \end{tabular}}
\end{minipage}
\end{table}
\vspace{-1em}
\subsubsection{Comparison of Different Multi-task Frameworks.}
To investigate the effectiveness of different multi-task learning frameworks, we compared Shared Bottom, MoE \cite{shazeer2017outrageously}, MMoE \cite{ma2018modeling}, and CGC \cite{dai2024gated}-based Multi-task DFNet under the same experimental settings. As shown in Table \ref{multitask}, the CGC-based Multi-task DFNet consistently outperforms the other frameworks across all evaluation metrics.
\vspace{-1em}

\begin{table}[]
\centering
\caption{Impact of multi-task frameworks on MIT-BIH/PTB performance.}\label{multitask}
\resizebox{0.6\textwidth}{!}{
\begin{tabular}{l|llll|llll}
\hline
 & \multicolumn{4}{c|}{MIT-BIH} & \multicolumn{4}{c}{PTB} \\ \cline{2-9}
 & Acc(\%) & Prec(\%) & Rec(\%) & F1(\%) & Acc(\%) & Prec(\%) & Rec(\%) & F1(\%) \\ \hline

SharedBottom & 97.33 & 94.84 & 85.79 & 89.68 & 99.67 & 98.33 & 97.87 & 98.09 \\
MoE          & 97.75 & 94.06 & 88.61 & 91.10 & 99.65 & 98.62 & 97.63 & 98.27 \\
MMoE         & 97.70 & 95.12 & 88.57 & 91.53 & 99.69 & 98.63 & 98.30 & 98.46 \\ \hline

\textbf{CGC} &
\textbf{99.01} & \textbf{98.02} & \textbf{91.11} & \textbf{94.22} &
\textbf{99.86} & \textbf{99.09} & \textbf{98.62} & \textbf{98.83} \\ \hline

\end{tabular}}
\end{table}

\vspace{-1em}
\subsubsection{Effect of GRU-Diffusion.}
\begin{figure}[htbp]
  \centering
  \includegraphics[width=\linewidth]{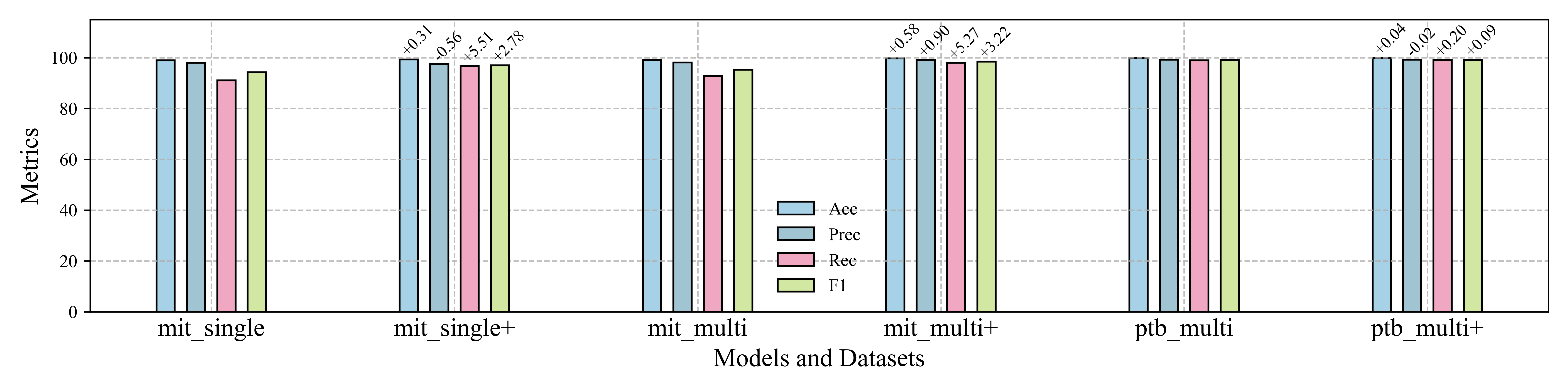}
  \caption{Effect of GRU-Diffusion augmentation on metrics (Single vs. Single+, Multi vs. Multi+).}\label{augment}
\end{figure}

We further evaluated the impact of GRU-augmented Diffusion on model performance. The detailed results are presented in Figure \ref{augment}.
For the MIT-BIH dataset, GRU-Diffusion was applied to augment the training data by generating synthetic ECG signals for underrepresented classes. As expected, Multi-task DFNet + GRU-Diffusion showed significant improvements in classification metrics compared to its counterpart without data augmentation.

Interestingly, although data augmentation was only applied to the MIT-BIH dataset, we observed that Multi-task DFNet + GRU-Diffusion also achieved improved performance on the PTB dataset. This surprising result suggests that incorporating GRU-Diffusion enhances the model's generalization capability across datasets, likely due to the improved feature representations learned from more diverse temporal patterns.
\vspace{-1em}
\vspace{-1em}

\subsection{Discussion}
Although GRU-augmented Diffusion improves ECG signal generation, it comes with the challenges of a larger model size and a slower generation process. However, given that we only generated a small amount of data, the additional computational cost is acceptable, especially considering the significant improvement in classification performance. The ability of GRU-Diffusion to address class imbalance and improve generalization across datasets, even without direct augmentation on the PTB dataset, further justifies its inclusion in our model for ECG classification tasks.
\vspace{-1em}
\section{Conclusion and Future work}
This paper proposed Multi-task DFNet, a lightweight and efficient framework for comprehensive ECG classification across the MIT-BIH and PTB datasets. By combining a CGC-based multi-task structure with GRU-augmented Diffusion for data augmentation, the model effectively addresses both the challenges of resource constraints and class imbalance. Extensive experiments demonstrate that Multi-task DFNet achieves superior performance with significantly fewer parameters and lower computational cost than traditional and lightweight baselines. These results highlight the potential of the proposed approach for real-time ECG analysis on edge devices.

However, MIT-BIH and PTB cover disjoint sets of diagnoses, which limits our ability to train a truly unified, multi‐label model on a single dataset. In future work, we will incorporate additional ECG repositories with overlapping and complementary annotations to enable one-dataset, multi-annotation learning.

%
%
%
%

\printbibliography
\newpage
\appendix
\noindent             
{\bfseries Appendix\par}  
\addcontentsline{toc}{section}{Appendix}  
\vspace{-2em}
\begin{table}[H]
\centering
\caption{Thorough Quantitative Evaluation of proposed Models}\label{bigtab}
\small
\resizebox{0.8\textwidth}{!}{
\begin{tabular}{llllllllll}
\hline
\multicolumn{10}{c}{\textbf{MITBIH Arrhythmia Database}}                                                                                                                                                              \\ \hline
\textbf{Single} & Acc             & Prec             & Rec             & \multicolumn{1}{l|}{F1}              & \textbf{Single+}     & Acc             & Pre                       & Rec             & F1              \\ \hline
NOR             & 0.9923          & 0.9890          & 0.9987          & \multicolumn{1}{l|}{0.9938}          & NOR                  & 0.9943          & 0.9922                    & 0.9986          & 0.9954          \\
LBB             & 1.0000          & 1.0000          & 1.0000          & \multicolumn{1}{l|}{1.0000}          & LBB                  & 1.0000          & 1.0000                    & 1.0000          & 1.0000          \\
PB              & 0.9997          & 0.9972          & 0.9981          & \multicolumn{1}{l|}{0.9976}          & PB                   & 0.9999          & 0.9991                    & 0.9991          & 0.9991          \\
FPN             & 0.9995          & 0.9902          & 0.9352          & \multicolumn{1}{l|}{0.9619}          & FPN                  & 0.9998          & 1.0000                    & 0.9630          & 0.9811          \\
RBB             & 0.9996          & 0.9983          & 0.9958          & \multicolumn{1}{l|}{0.9971}          & RBB                  & 0.9999          & 1.0000                    & 0.9983          & 0.9992          \\
APB             & 0.9966          & 0.9891          & 0.8771          & \multicolumn{1}{l|}{0.9298}          & APB                  & 0.9966          & 0.9891                    & 0.8771          & 0.9298          \\
\textbf{AAP}    & 0.9992          & 1.0000          & 0.7234          & \multicolumn{1}{l|}{0.8395}          & \textbf{AAP}         & \textbf{0.9994} & \textbf{0.8936}           & \textbf{0.8936} & \textbf{0.8936} \\
PVC             & 0.9967          & 0.9854          & 0.9864          & \multicolumn{1}{l|}{0.9859}          & PVC                  & 0.9978          & 0.9932                    & 0.9880          & 0.9906          \\
\textbf{FVN}    & 0.9977          & 0.9388          & 0.7419          & \multicolumn{1}{l|}{0.8288}          & \textbf{FVN}         & \textbf{0.9993} & \textbf{0.9748}           & \textbf{0.9355} & \textbf{0.9547} \\
\textbf{VFW}    & 0.9999          & 1.0000          & 0.9286          & \multicolumn{1}{l|}{0.9630}          & \textbf{VFW}         & \textbf{0.9999} & \textbf{0.9643}           & \textbf{0.9643} & \textbf{0.9643} \\
\textbf{JEB}    & 0.9991          & 0.8750          & 0.7955          & \multicolumn{1}{l|}{0.8333}          & \textbf{JEB}         & \textbf{0.9998} & \textbf{0.9348}           & \textbf{0.9773} & \textbf{0.9556} \\
\textbf{VEB}    & 0.9999          & 1.0000          & 0.9524          & \multicolumn{1}{l|}{0.9756}          & \textbf{VEB}         & \textbf{0.9999} & \textbf{0.9545}           & \textbf{1.0000} & \textbf{0.9767} \\ \hline
\textbf{Avg}    & \textbf{0.9901} & \textbf{0.9802} & \textbf{0.9111} & \multicolumn{1}{l|}{\textbf{0.9422}} & \textbf{Avg}         & \textbf{0.9932} & \textbf{0.9746}           & \textbf{0.9662} & \textbf{0.9700} \\ \hline
\textbf{Multi}  & Acc             & Pre             & Rec             & \multicolumn{1}{l|}{F1}              & \textbf{Multi+}      & Acc             & Pre                       & Rec             & F1              \\ \hline
NOR             & 0.9928          & 0.9903          & 0.9982          & \multicolumn{1}{l|}{0.9943}          & NOR                  & 0.9975          & 0.9967                    & 0.9992          & 0.9980          \\
LBB             & 1.0000          & 1.0000          & 1.0000          & \multicolumn{1}{l|}{1.0000}          & LBB                  & 1.0000          & 1.0000                    & 1.0000          & 1.0000          \\
PB              & 0.9999          & 0.9991          & 0.9991          & \multicolumn{1}{l|}{0.9991}          & PB                   & 0.9999          & 1.0000                    & 0.9991          & 0.9995          \\
FPN             & 0.9997          & 1.0000          & 0.9537          & \multicolumn{1}{l|}{0.9763}          & FPN                  & 0.9998          & 0.9907                    & 0.9815          & 0.9860          \\
RBB             & 0.9996          & 0.9975          & 0.9975          & \multicolumn{1}{l|}{0.9975}          & RBB                  & 1.0000          & 1.0000                    & 1.0000          & 1.0000          \\
APB             & 0.9969          & 0.9841          & 0.8940          & \multicolumn{1}{l|}{0.9369}          & APB                  & 0.9990          & 0.9901                    & 0.9687          & 0.9793          \\
\textbf{AAP}    & 0.9994          & 0.9750          & 0.8298          & \multicolumn{1}{l|}{0.8966}          & \textbf{AAP}         & \textbf{0.9996} & \textbf{0.9767}           & \textbf{0.8936} & \textbf{0.9333} \\
PVC             & 0.9973          & 0.9890          & 0.9880          & \multicolumn{1}{l|}{0.9885}          & PVC                  & 0.9990          & 0.9974                    & 0.9943          & 0.9958          \\
\textbf{FVN}    & 0.9980          & 0.9423          & 0.7903          & \multicolumn{1}{l|}{0.8596}          & \textbf{FVN}         & \textbf{0.9998} & \textbf{1.0000}           & \textbf{0.9677} & \textbf{0.9836} \\
\textbf{VFW}    & 0.9999          & 1.0000          & 0.9286          & \multicolumn{1}{l|}{0.9630}          & \textbf{VFW}         & \textbf{1.0000} & \textbf{1.0000}           & \textbf{1.0000} & \textbf{1.0000} \\
\textbf{JEB}    & 0.9992          & 0.8974          & 0.7955          & \multicolumn{1}{l|}{0.8434}          & \textbf{JEB}         & \textbf{0.9998} & \textbf{0.9767}           & \textbf{0.9545} & \textbf{0.9655} \\
\textbf{VEB}    & 0.9999          & 1.0000          & 0.9524          & \multicolumn{1}{l|}{0.9756}          & \textbf{VEB}         & \textbf{0.9999} & \textbf{0.9545}           & \textbf{1.0000} & \textbf{0.9767} \\ \hline
\textbf{Avg}    & \textbf{0.9914} & \textbf{0.9812} & \textbf{0.9272} & \multicolumn{1}{l|}{\textbf{0.9526}} & \textbf{Avg}         & \textbf{0.9972} & \textbf{0.9902}           & \textbf{0.9799} & \textbf{0.9848} \\ \hline
\multicolumn{10}{c}{\textbf{PTB Diagnostic ECG Database}}                                                                                                                                                                            \\ \hline
\textbf{Single} & Acc             & Pre             & Rec             & \multicolumn{1}{l|}{F1}              & \multicolumn{1}{c}{} &                 &                           &                 &                 \\ \hline
Bun             & 0.9999          & 1.0000          & 0.9971          & \multicolumn{1}{l|}{0.9985}          &                      &                 &                           &                 &                 \\
Car             & 0.9999          & 0.9973          & 1.0000          & \multicolumn{1}{l|}{0.9986}          &                      &                 &                           &                 &                 \\
Dys             & 0.9998          & 1.0000          & 0.9916          & \multicolumn{1}{l|}{0.9958}          &                      &                 &                           &                 &                 \\
Hea             & 0.9999          & 1.0000          & 0.9994          & \multicolumn{1}{l|}{0.9997}          &                      &                 &                           &                 &                 \\
Hyp             & 0.9990          & 0.9451          & 0.9873          & \multicolumn{1}{l|}{0.9657}          &                      &                 & \multicolumn{1}{c}{-----} &                 &                 \\
Mis             & 1.0000          & 1.0000          & 1.0000          & \multicolumn{1}{l|}{1.0000}          &                      &                 &                           &                 &                 \\
MI              & 0.9997          & 0.9996          & 1.0000          & \multicolumn{1}{l|}{0.9998}          &                      &                 &                           &                 &                 \\
Myo             & 0.9990          & 0.9759          & 0.9000          & \multicolumn{1}{l|}{0.9364}          &                      &                 &                           &                 &                 \\
Val             & 1.0000          & 1.0000          & 1.0000          & \multicolumn{1}{l|}{1.0000}          &                      &                 &                           &                 &                 \\ \hline
\textbf{Avg}    & \textbf{0.9986} & \textbf{0.9909} & \textbf{0.9862} & \multicolumn{1}{l|}{\textbf{0.9883}} &                      &                 &                           &                 &                 \\ \hline
\textbf{Multi}  & Acc             & Pre             & Rec             & \multicolumn{1}{l|}{F1}              & \textbf{Multi+}      & Acc             & Pre                       & Rec             & F1              \\ \hline
Bun             & 1.0000          & 1.0000          & 1.0000          & \multicolumn{1}{l|}{1.0000}          & Bun                  & 1.0000          & 1.0000                    & 1.0000          & 1.0000          \\
Car             & 0.9997          & 1.0000          & 0.9918          & \multicolumn{1}{l|}{0.9959}          & Car                  & 1.0000          & 1.0000                    & 1.0000          & 1.0000          \\
Dys             & 0.9997          & 1.0000          & 0.9874          & \multicolumn{1}{l|}{0.9937}          & Dys                  & 0.9997          & 1.0000                    & 0.9874          & 0.9937          \\
Hea             & 0.9998          & 1.0000          & 0.9989          & \multicolumn{1}{l|}{0.9994}          & Hea                  & 0.9998          & 0.9994                    & 0.9994          & 0.9994          \\
Hyp             & 0.9993          & 0.9686          & 0.9809          & \multicolumn{1}{l|}{0.9747}          & Hyp                  & 0.9994          & 0.9870                    & 0.9682          & 0.9775          \\
Mis             & 1.0000          & 1.0000          & 1.0000          & \multicolumn{1}{l|}{1.0000}          & Mis                  & 1.0000          & 1.0000                    & 1.0000          & 1.0000          \\
MI              & 0.9991          & 0.9989          & 0.9999          & \multicolumn{1}{l|}{0.9994}          & MI                   & 0.9996          & 0.9995                    & 1.0000          & 0.9997          \\
Myo             & 0.9993          & 0.9659          & 0.9444          & \multicolumn{1}{l|}{0.9551}          & Myo                  & 0.9993          & 0.9457                    & 0.9667          & 0.9560          \\
Val             & 1.0000          & 1.0000          & 1.0000          & \multicolumn{1}{l|}{1.0000}          & Val                  & 1.0000          & 1.0000                    & 1.0000          & 1.0000          \\ \hline
\textbf{Avg}    & \textbf{0.9985} & \textbf{0.9926} & \textbf{0.9893} & \multicolumn{1}{l|}{\textbf{0.9909}} & \textbf{Avg}         & \textbf{0.9989} & \textbf{0.9924}           & \textbf{0.9913} & \textbf{0.9918} \\ \hline
\end{tabular}}
\end{table}

\end{document}